\documentclass[pdflatex,sn-vancouver-num]{sn-jnl}% Vancouver Numbered Reference Style
\usepackage{graphicx}%
\usepackage{float}
\usepackage{svg}
\usepackage{multirow}%
\usepackage{amsmath,amssymb,amsfonts}%
\usepackage{amsthm}%
\usepackage{mathrsfs}%
\usepackage[title]{appendix}%
\usepackage{xcolor}%
\usepackage{textcomp}%
\usepackage{manyfoot}%
\usepackage{booktabs}%
\usepackage{enumitem}
\usepackage{algorithm}%
\usepackage{algorithmicx}%
\usepackage{algpseudocode}%
\usepackage{listings}%

\usepackage{caption}
\usepackage{subcaption}
\usepackage{color}
\usepackage{soul}
\sethlcolor{white} % removes edits yellow highlighting

\theoremstyle{thmstyleone}%
\theoremstyle{thmstyletwo}%

\theoremstyle{thmstylethree}%

\begin{document}

\title[An Economic Analysis of DNA-based Data Storage Systems]{An Economic Analysis of DNA-based Data Storage Systems}


\author*[1]{\fnm{Alex} \sur{El-Shaikh}}\email{a.elshaikh@imperial.ac.uk}

\author[2]{\fnm{Bernhard} \sur{Seeger}}\email{seeger@mathematik.uni-marburg.de}

\author[1]{\fnm{Thomas} \sur{Heinis}}\email{t.heinis@imperial.ac.uk}

\affil[1]{\orgdiv{Department of Computing}, \orgname{Imperial College London}, \orgaddress{\street{180 Queen's Gate}, \postcode{SW7 2RH}, \country{United Kingdom}}}

\affil[2]{\orgdiv{Department of Mathematics and Computer Science}, \orgname{University of Marburg}, \orgaddress{\street{Hans-Meerwein-Str. 6}, \city{Marburg}, \postcode{35043}, \state{Hesse}, \country{Germany}}}

%\affil[3]{\orgdiv{Department}, \orgname{Organization}, \orgaddress{\street{Street}, \city{City}, \postcode{610101}, \state{State}, \country{Country}}}


\abstract{Deoxyribonucleic acid (DNA) remains stable for millennia without degradation, can store information at densities orders of magnitude higher than current technologies, and is environmentally friendly due to its low energy requirement. While these advantages make it a promising candidate for storing archival data, DNA storage systems are constrained by high costs, particularly those associated with DNA synthesis. In this paper, we present a comprehensive cost model for calculating the cost of archiving data using various storage systems. Leveraging our model, we conduct a cost analysis of DNA storage versus leading magnetic tape and cloud-based archival storage services. Furthermore, this model can be customised with various parameters to reflect future cost declines and other cost-relevant trends. Our results indicate that, under current cost trajectories, DNA storage costs must fall by eight to nine orders of magnitude to become economically competitive with today's data archival technologies. Moreover, we explore multiple ``what‑if'' scenarios over the coming decades, quantifying the rate of cost decline required to close the competitiveness gap relative to tape and cloud archives. Our findings underscore the critical need for accelerated innovation and investment in DNA synthesis technologies to reduce their cost and transform DNA storage into a practical archival solution.}


\maketitle

\section{Introduction}\label{sec1}

Driven by rapid growth in cloud services, artificial intelligence (AI), high-resolution multimedia, Internet of Things (IoT) deployments, and large-scale scientific experiments, global data production is increasing exponentially and at an ever-increasing pace \cite{10.1145/3696672, ma2020big}. This trend poses unprecedented challenges for traditional storage technologies, as current storage capacity supply is already outpaced by the demand \cite{li2020can}. Additionally, the growing demand for long-term archival storage drives up costs for traditional media, such as disks and tapes, which must be replaced every 5-30 years to prevent data loss. This recurring cycle of media replacements puts additional strain on the storage infrastructure.

In the face of these challenges, Deoxyribonucleic acid (DNA) storage emerges as a promising alternative \cite{li2023dp}. It can store an astounding 455 exabytes of data per gram, around six orders of magnitude more than traditional storage \cite{XU2025624}. Furthermore, DNA is exceptionally durable, capable of preserving information for thousands of years with minimal energy requirements under suitable conditions, making it both environmentally friendly and sustainable \cite{meiser2022synthetic, panda2018dna}. While current DNA storage technologies are limited by slow and costly read/write processes, such constraints are less critical for managing archival data. Thus, given that most of the data stored today is considered cold \cite{10.1145/3696672}, i.e., not accessed frequently, DNA storage is particularly well suited for cold data management \cite{appuswamy2019oligoarchive}.

Despite the advantages mentioned above, implementing DNA-based storage suffers from drawbacks. First, the associated costs for DNA reading (sequencing) and writing (synthesis) today are substantially higher than for traditional technologies. While DNA sequencing is approximately $1{,}000$ times more expensive than reading data from tape or disk, DNA synthesis is eight to ten orders of magnitude (i.e., 100 million to 1 billion times) more costly than writing to traditional media. Consequently, DNA synthesis remains the dominant cost driver in DNA storage systems.
%Second, current DNA synthesis and sequencing technologies are prone to errors, necessitating increased redundancy and the use of additional error-correcting mechanisms \cite{10.1145/3708997, Mortuza2023}.
Second, DNA sequencing and synthesis are slower than for tape or disk \cite{cao2024efficient, Zhou2024}. Unlike traditional storage systems, DNA storage is not a drop-in technology; instead, it depends on wet-lab equipment and trained personnel, making its operation less straightforward.

Estimating the total cost of storing data with a given technology requires accounting for the costs of writing, reading, and maintaining the storage media over the retention period.
%Traditionally, cost-driven decisions in storage systems have been guided by heuristics such as the 5-minute rule, introduced in 1987 \cite{gray19875}. This rule suggested that a data page should be kept in memory if it is accessed more frequently than once every five minutes, based on the cost trade-off between RAM and disk storage in 1987. Since then, this figure has been updated multiple times, and the method has been extended to other storage layers, including SSDs, tape, and cloud \cite{appuswamy2017five, graefe2007five, gray1997five}. Notably, the 5-minute rule was not proposed to include the cost of storing data over a specific period. Instead, it focuses on the data size and its access frequency, making it less ideal for estimating total storage costs.
This paper introduces a parameterised cost model to estimate the resulting cost for storing data using a specific storage technology. It breaks down the total cost into the three components as mentioned above: (i) write cost, (ii) read cost, and (iii) running and maintenance cost. Core input parameters include the number of objects, object size, read frequency, and storage duration. Additionally, the model is designed to include expenses for storing data over a given storage duration. Thus, it also accounts for storage media degradation by incorporating the cost of potential media migration required over the retention period.

Furthermore, this paper conducts a cost analysis of DNA storage in comparison to cloud storage services, such as Amazon S3 Glacier Deep Archive and Azure Blob Archive, and to tape on-premise storage. 
%Cloud services provide transparent pricing for end users, whereas on-premises tape requires self-management, adding operational complexity but typically offering a lower overall cost.
Among our results, we reveal that the currently exceptionally high DNA synthesis cost dominates the total cost for DNA storage, leading to the formulation of the following four key hypotheses \textbf{H1-H4}:
\begin{enumerate}[label=\textbf{H\arabic*}]
    \item The high cost of DNA synthesis remains the dominant barrier preventing economic adoption of DNA storage.
    
    \item Achieving economic competitiveness with traditional archival media (e.g., tape) will require breakthroughs in DNA synthesis technology that surpass historical exponential cost reduction trends.

    \item Improvements in logical encoding density (bits per DNA base) alone provide limited economic benefits, insufficient to overcome the high costs associated with DNA synthesis.

    %\item DNA storage's exceptional longevity ($> 1000$ years) provides substantial economic potential, particularly for ultra-long-term archival applications.

    \item The unique characteristics of DNA storage (extreme longevity, ultra-high density, passive energy-free maintenance, and superior sustainability) position it favorably for entirely new application classes that conventional storage methods cannot serve.

    %\item DNA storage can substantially reduce data loss as current storage technologies continue to develop.

\end{enumerate}
These hypotheses guide our subsequent analysis and evaluation. In \textbf{Methods}, we introduce our cost model to quantitatively assess the economic viability of DNA storage under various technological and workload scenarios. The \textbf{Results} section explores the cost trade-offs, sensitivity to technological improvements, and identifies conditions under which DNA storage could become a competitive or superior option compared to traditional archival storage technologies.
%Moreover, trends in DNA sequencing costs indicate that reading data from DNA may close the gap with tape by 2030, historical trends in the declining cost of DNA synthesis suggest that cost parity with tape storage could take as long as 300 years. Therefore, increased improvement in DNA synthesis beyond what the current trends indicate is necessary.

\section{Results}\label{sec2}

\subsection{Overview}
We evaluate storage costs using DNA storage \cite{wetterstrand2023dnacosts, lin2022enable, carlson2022dna, kosuri2014large, Carr2009}, cloud archival services such as Amazon S3 Glacier Deep Archive \cite{aws2025} and Azure Blob Archive \cite{azure2025}, and tape on-premise solutions \cite{fujifilm2025tcotool, 10683720}. Cloud storage services provide transparent pricing for end users, whereas on-premises tape requires self-management, adding operational complexity but typically offering a lower overall cost. To estimate DNA's costs in the future, we used an exponential fit for each of DNA synthesis and DNA sequencing, computed from historical cost data. \hl{Our extrapolations are not intended as a guarantee that past trends will persist indefinitely; they are a scenario to translate how fast costs must fall into an economic implication for DNA storage. Our projections are used as a baseline, and we test sensitivity to slower post-trend decline rates and/or cost floors to reflect potential physical, chemical, and economic limits.} The explicit costs for writing, reading, and maintenance for DNA and the other storage options are further presented in \textbf{Methods}.

We refer to the Amazon S3 Glacier Deep Archive as Amazon Deep Archive, which is set as the default tape/non-DNA storage solution, as the other storage solutions, such as Azure Blob Archive and tape on-premise, are similar in costs and storage operations.

\subsection{Parameters Setting}
Based on recent historical trends, the cost of tape-based storage is set to drop by a fixed value of 10\% every year \cite{Doricchi2022, 10.1063/1.5007621, 10683720, national2024rapid}. Moreover, the default lifetime for tape is set to 30 years \cite{Mortuza2023}, whereas DNA's default lifetime is set to 1000 years \cite{RAZA2023108155}. The other parameters that are not related to the storage technology are set to their defaults as follows: the number of data objects $(n=1{,}000)$, the number of read objects per year $(k=10)$, the object size in MB $(\mathit{obj\_size}=1{,}000)$, and the storage period in years $(d=100)$. These settings yield a 1~TB archive, a realistic collection-level size in institutional and enterprise archives aligned with common 1--5~TB project quotas and conservative against LTO-9 media capacity (18~TB), yet larger than current DNA-storage demonstrations \cite{ibmLTO9cap, Organick2018, oscProj1to5TB, ubCCR1TB}.

In terms of access patterns, studies show that the vast majority of archival data remains untouched after being written: the Storage Networking Industry Association (SNIA) reports that up to 90\% of enterprise network-attached storage (NAS) data is never accessed again \cite{snia2016role}, and that cold data archives are explicitly designed for ``write once, read rarely'' usage \cite{eurekalert2024tape}. Industry analysis further indicates that up to 80\% of enterprise data is considered cold with infrequent access spanning minutes to years \cite{fmr2024sustainable}. In line with these observations, we assume that 10~GB of the 1~TB archive is read every year (i.e., $\approx$1\% annual access), which is consistent with large-scale data archive behaviour \cite{berman2022xrootd}.

A byte of data can be stored in four DNA bases, giving a theoretical density of two bits per base. In reality, biochemical limits such as homopolymers, GC balance, and secondary structures require extra redundancy, reducing efficiency \cite{Lin2025, IJAIN1747}. In our model, we assume two bits per base, in line with recent approaches that nearly reach this rate \cite{10.1145/3626233}. Some methods even exceed this rate by applying compression \cite{https://doi.org/10.1002/smtd.202101335, Zheng2024}. However, changing the bit rate has a negligible effect on overall costs compared to the significantly higher baseline cost of DNA storage. Finally, all monetary expenses provided in our experiments are shown in U.S. dollars.

\subsection{Evaluation}
\begin{figure}[H]
    \centering
    \includegraphics[width=0.6\textwidth]{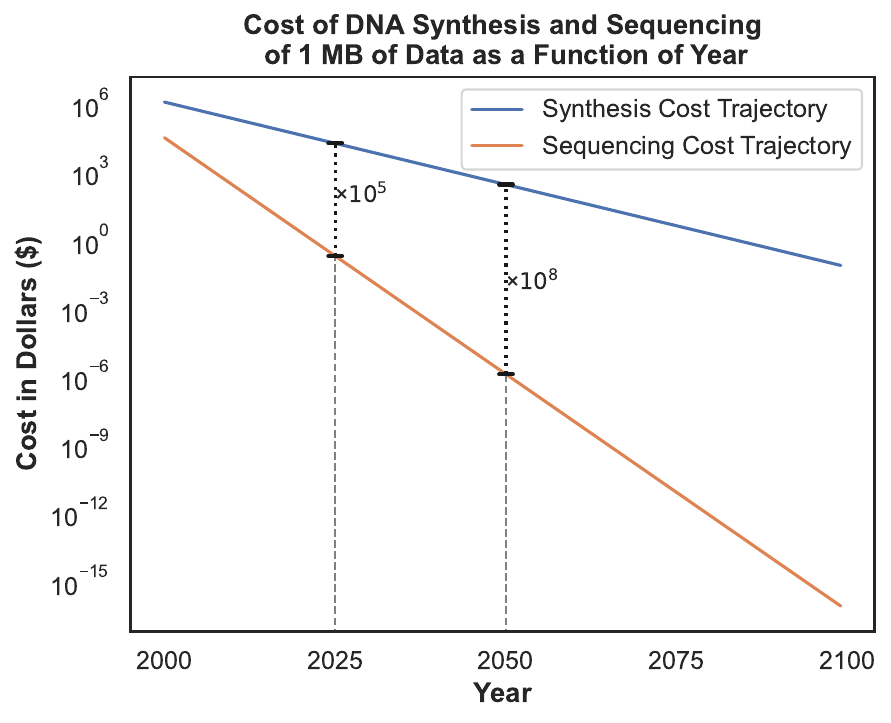}
    \caption{Comparing the cost of DNA synthesis and sequencing as a function of year.} \label{h1.1}
\end{figure}

\autoref{h1.1} depicts the cost of DNA synthesis and sequencing of 1~MB of data over time. By fitting exponential curves to historical pricing data, we project these trends forward to forecast future cost trajectories. Today, DNA synthesis remains approximately five orders of magnitude more expensive than DNA sequencing. Put differently, writing 1~MB of data costs as much as reading 10~GB of data using DNA.
The historical annual cost drop rate, i.e., the percentage drop per year, is estimated at roughly $47.9\%$ for DNA sequencing and $16.7\%$ for DNA synthesis. As a result, DNA synthesis is currently the dominant cost driver in DNA storage systems, accounting for more than $99.9\%$ of the total cost, even under read-intensive scenarios. This cost disparity is projected to widen further. For example, by 2050, the gap between DNA synthesis and sequencing may increase from five to eight orders of magnitude, reinforcing DNA synthesis as the bottleneck in terms of cost.

\begin{figure}[H]
	\centering
	\begin{subfigure}[t]{0.79\textwidth}
		\centering
		\includegraphics[width=\textwidth]{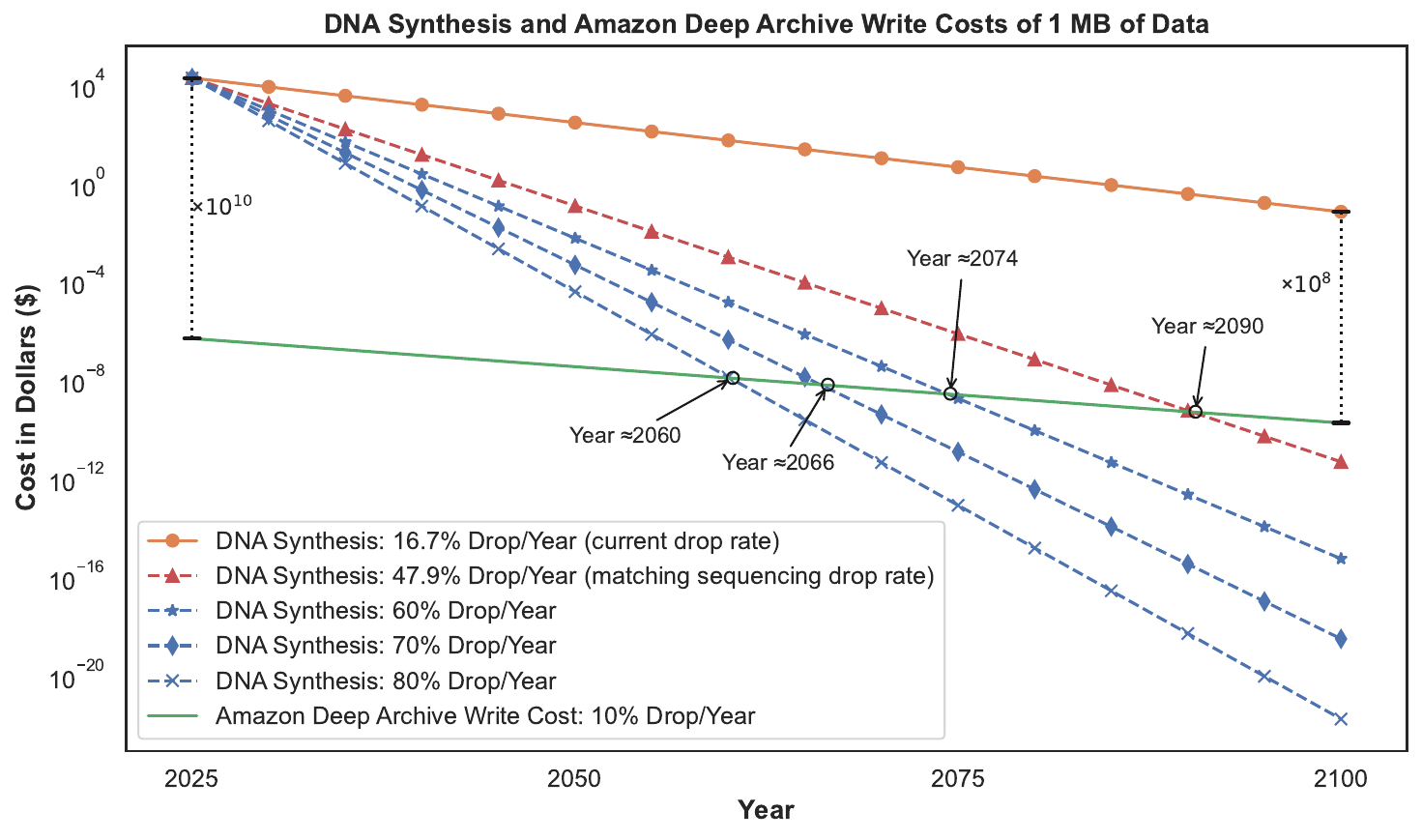}
		\caption{}
		\label{h6.1}
	\end{subfigure}
	\hspace{1em}%\hfill
	\begin{subfigure}[t]{0.79\textwidth}
		\centering
		\includegraphics[width=\textwidth]{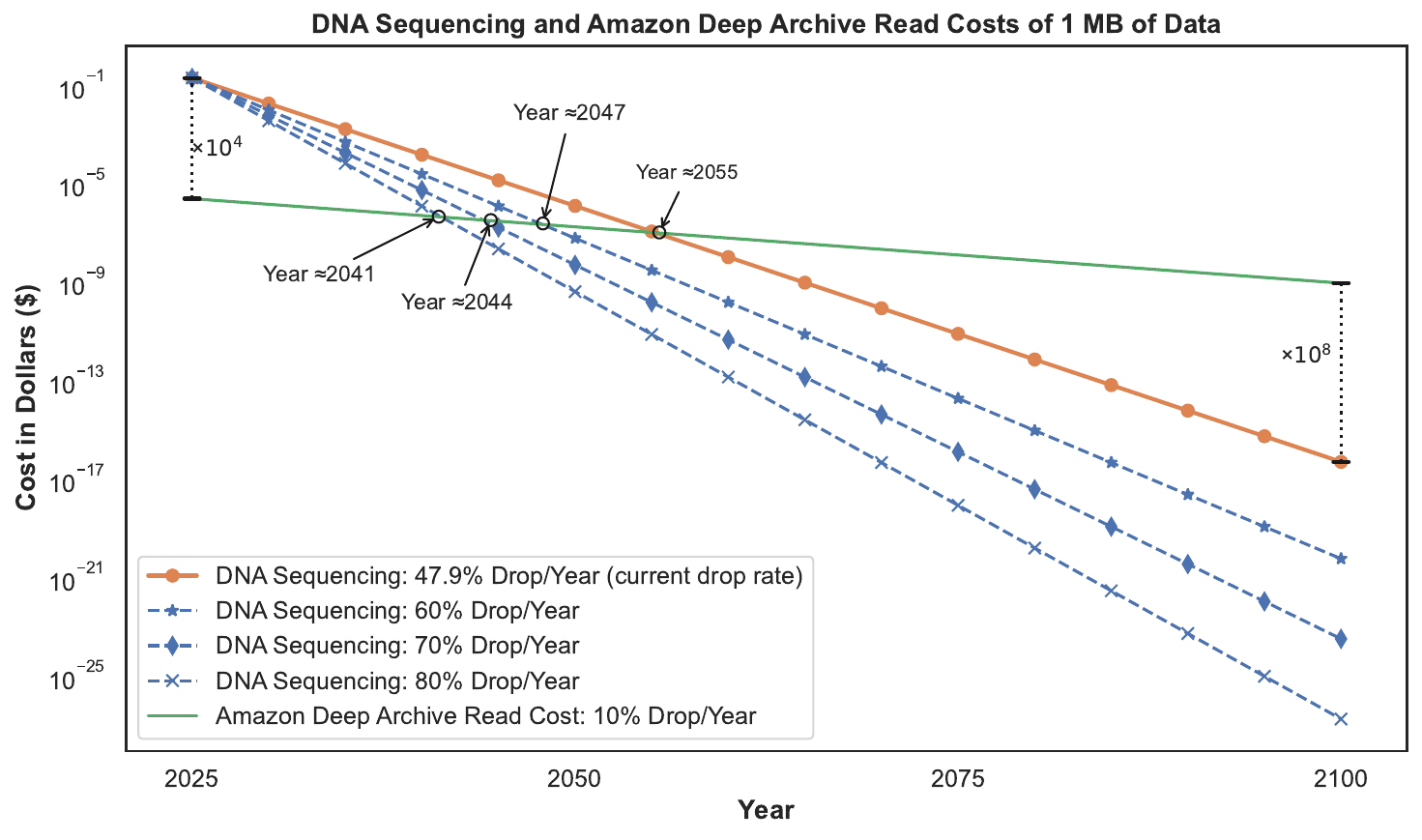}
		\caption{}
		\label{h6.2}
	\end{subfigure}
    \hspace{1em}%\hfill
	\begin{subfigure}[t]{0.79\textwidth}
		\centering
		\includegraphics[width=\textwidth]{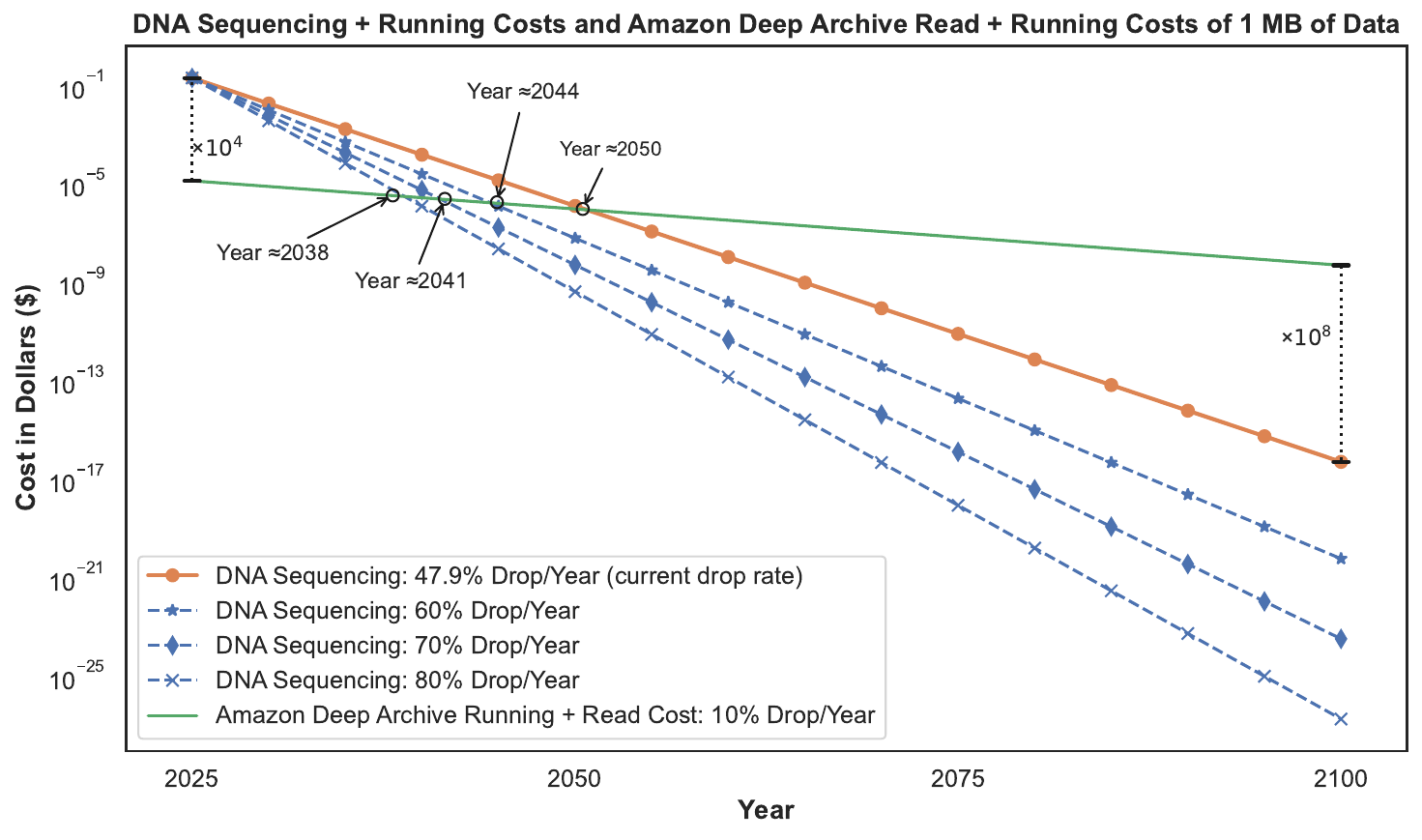}
		\caption{}
		\label{h6.3}
	\end{subfigure}
    \caption{Comparing the cost of DNA synthesis and sequencing to the write, running, and read cost using Amazon Deep Archive.}\label{h6.0}
\end{figure}

Today, writing data to traditional storage media is approximately ten orders of magnitude cheaper than writing it to DNA. As shown in \autoref{h6.1}, this cost gap is projected to narrow only slightly under the current trend, decreasing from ten to eight orders of magnitude by the end of this century. Assuming these trends continue without significant breakthroughs, write cost parity between DNA and traditional media is not expected to be reached until the 2300s.

However, if a technological breakthrough or other economic incentives were to accelerate the rate of cost reduction in DNA synthesis, the outlook could change significantly. For instance, if the annual drop rate in synthesis matched that of DNA sequencing (47.9\%) from 2025 onward, write cost parity with tape storage could be achieved as early as 2090. Under even more optimistic scenarios, e.g., assuming annual cost reductions of 60\%, 70\%, or 80\%, parity could be reached by 2074, 2066, and 2060, respectively. \textbf{Supplementary Figure 1} suggests similar estimations using tape on-premise and Azure Blob Storage.

In contrast, DNA sequencing is significantly cheaper than synthesis, yet it remains approximately five orders of magnitude more expensive than reading tape. Nevertheless, the cost of DNA sequencing has been declining much faster than synthesis in recent years, driven by advances in high-throughput platforms and automation, indicating a faster trajectory toward cost competitiveness with traditional media \cite{biology12070997, eren2022dna}. As illustrated in \autoref{h6.2}, DNA sequencing is projected to reach cost parity with reading tape by 2055 if the current trend continues. Assuming more optimistic scenarios, e.g., with annual cost drop rates of 60\%, 70\%, and 80\%, the parity could be achieved as early as 2047, 2044, and 2041. \autoref{h6.3} accounts for running and reading costs, resulting in cost parity even sooner, e.g., by the year 2050 following today's cost trends. Similar results are obtained when comparing Azure Blob Archive and tape on-premise, as presented in \textbf{Supplementary Figure 2}.

%Furthermore, the total cost of implementing and operating a data storage system depends on the workload characteristics. For instance, read-intensive workloads amplify the read-related costs. Additionally, the size of the data object ($\mathit{obj\_size}$) influences the resulting cost. For example, smaller objects result in more read operations for the same data volume than larger ones. The cost analysis presented in \autoref{h6.0} is based on writing and reading a single 1~MB data object. However, in real-world scenarios, the system parameters can vary.

\begin{figure}[H]
	\centering
	\begin{subfigure}[t]{0.75\textwidth}
		\centering
		\includegraphics[width=\textwidth]{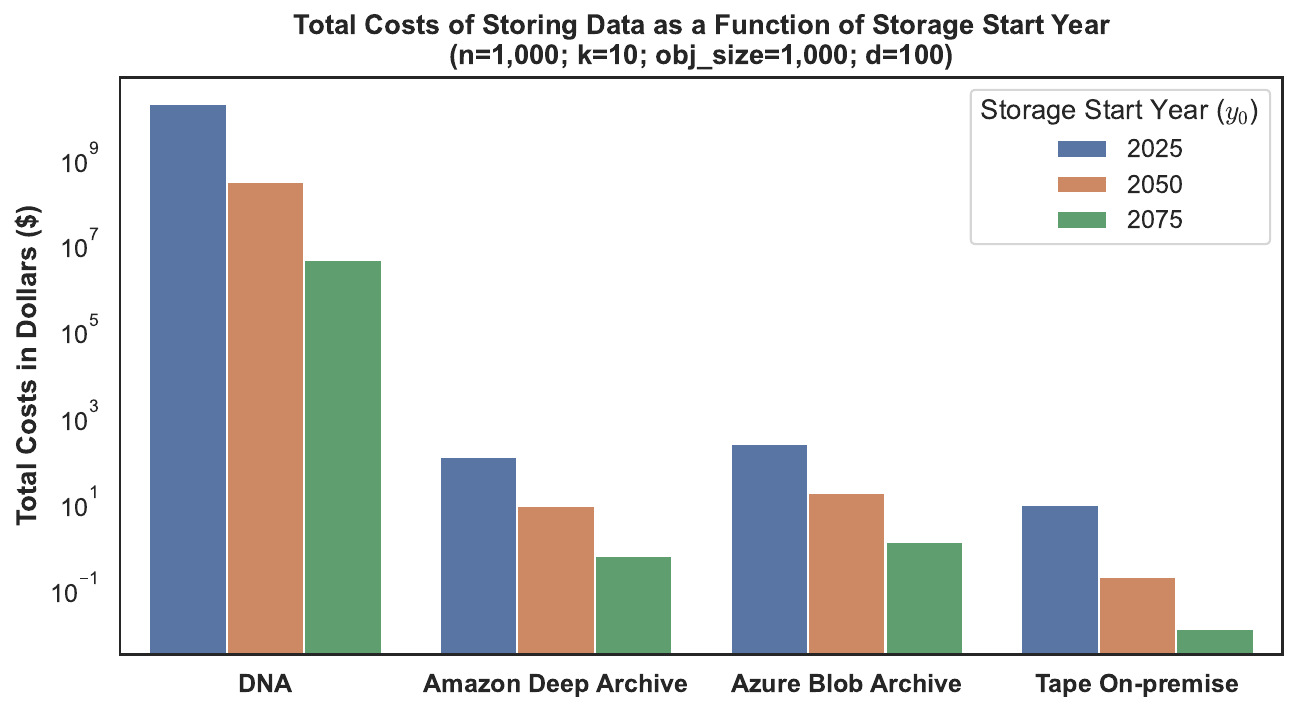}
		%\caption{The total cost of using DNA storage, Amazon Deep Archive, and Azure Blob Archive.}
        \caption{}
		\label{h1.2}
	\end{subfigure}
	\hspace{1em}%\hfill
	\begin{subfigure}[t]{0.75\textwidth}
		\centering
		\includegraphics[width=\textwidth]{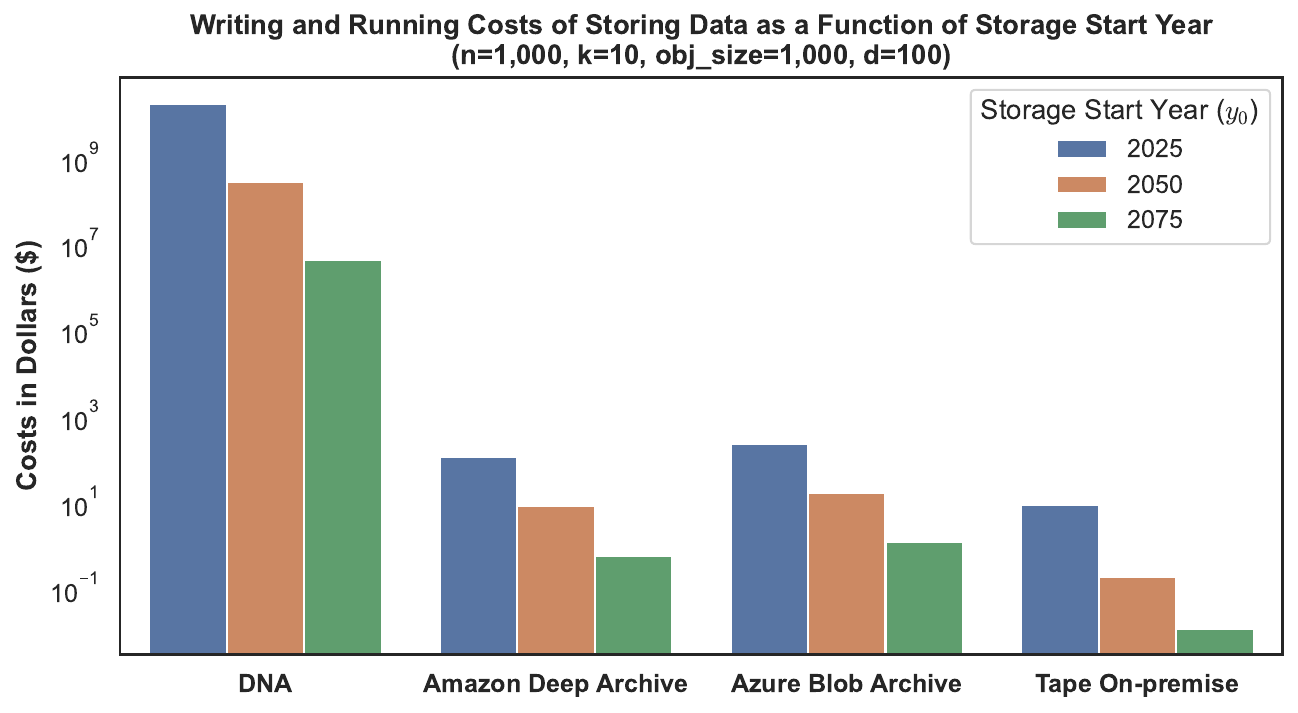}
		%\caption{Writing and running costs of using DNA storage, Amazon Deep Archive, and Azure Blob Archive.}
        \caption{}
		\label{h1.3}
	\end{subfigure}
    \hspace{1em}%\hfill
	\begin{subfigure}[t]{0.75\textwidth}
		\centering
		\includegraphics[width=\textwidth]{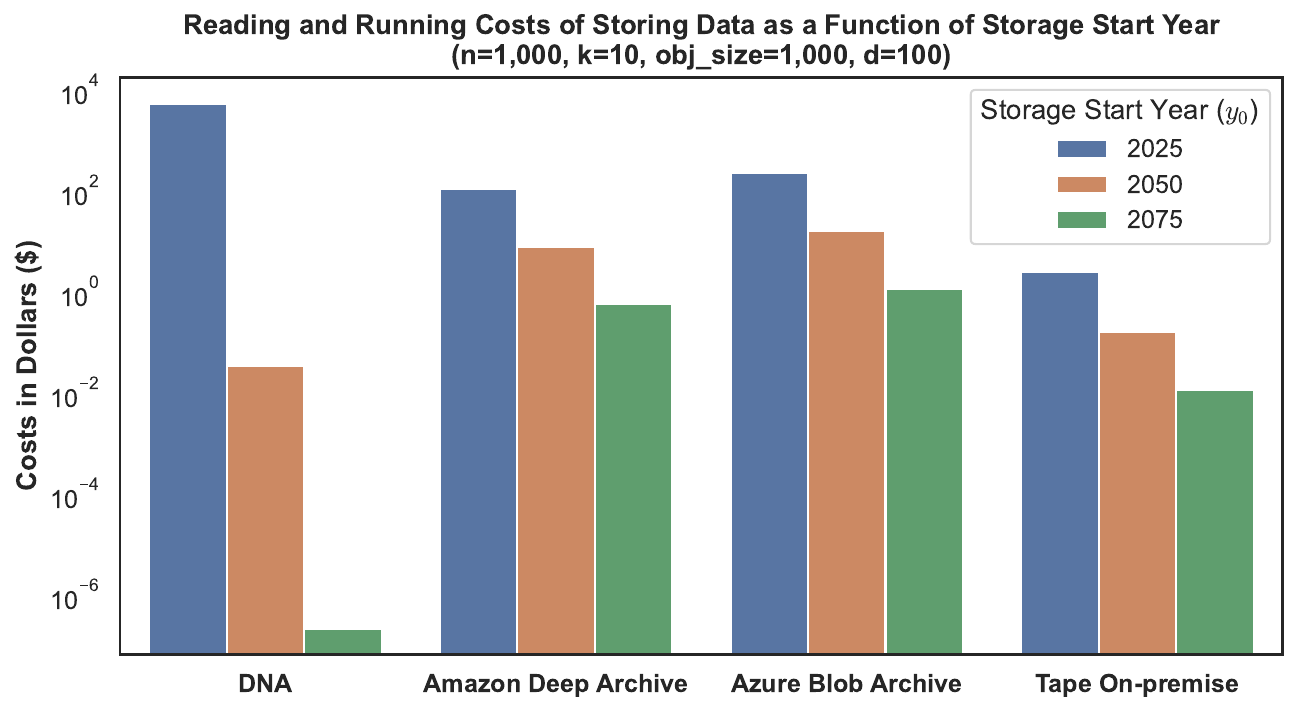}
		%\caption{Reading and running costs of using DNA storage, Amazon Deep Archive, and Azure Blob Archive.}
        \caption{}
		\label{h1.4}
	\end{subfigure}
    \caption{Comparing the different costs (write, read, run) of using DNA storage, Amazon Deep Archive, and Azure Blob Archive as a function of storage start year $y_0$.}\label{h1.2+3+4}
\end{figure}

The following \autoref{h1.2+3+4} compares the operating costs of the four storage systems (DNA, Amazon Deep Archive, Azure Blob Archive, and tape on-premise) as a function of the storage start year ($y_0$) using our default storage parameters setting. \autoref{h1.2} presents the total cost, \autoref{h1.3} isolates the write plus running cost, and \autoref{h1.4} isolates the read plus running cost. The near-perfect overlap of \autoref{h1.2} and \autoref{h1.3} demonstrates that write costs overwhelmingly drive total expenditure, with read costs being relatively insignificant.

Furthermore, \autoref{h1.4} shows that the cost of DNA sequencing could already be more cost-efficient than the cost of reading plus running tape storage by the year 2050.

Among the tape-based options, cost differences are relatively small, with tape on-premise being approximately two to three times less expensive than the cloud-based alternatives. 
%The reported costs account for all the necessary write, read, maintenance, and device replacement costs over the storage duration of $d=100$ years. The total costs are calculated using the cost model introduced in \textbf{Methods}.
Note that despite only $1\%$ of the total data being read every year, DNA storage remains prohibitively expensive, between seven and eight orders of magnitude higher under the given parameters. \textbf{Supplementary Figures 3-4} use different parameters to obtain similar results. Additionally, the overall cost is not significantly affected by the storage duration ($d$), as the cost for DNA synthesis and sequencing becomes exponentially less expensive over time, making data migration costs negligible.

\begin{figure}[H]
	\centering
	\begin{subfigure}[t]{0.82\textwidth}
		\centering
		\includegraphics[width=\textwidth]{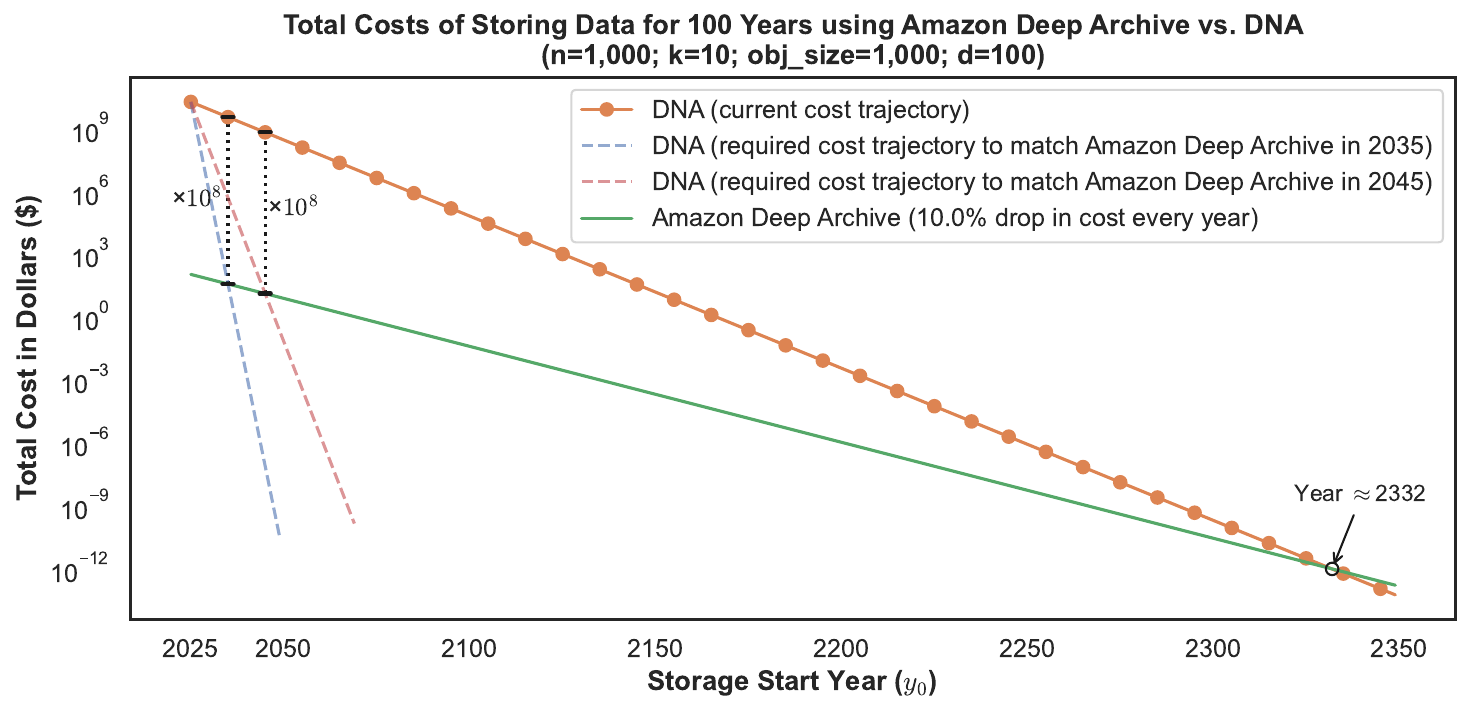}
		%\caption{The total cost trajectories for DNA storage vs. Amazon Deep Archive. The point at which DNA first reaches cost parity with Amazon is marked, and two optimistic DNA cost-decline scenarios for the years 2023 and 2045 with their respective crossover years are also shown.}
        \caption{}
		\label{h2.1}
	\end{subfigure}
	\hspace{1em}%\hfill
	\begin{subfigure}[t]{0.82\textwidth}
		\centering
		\includegraphics[width=\textwidth]{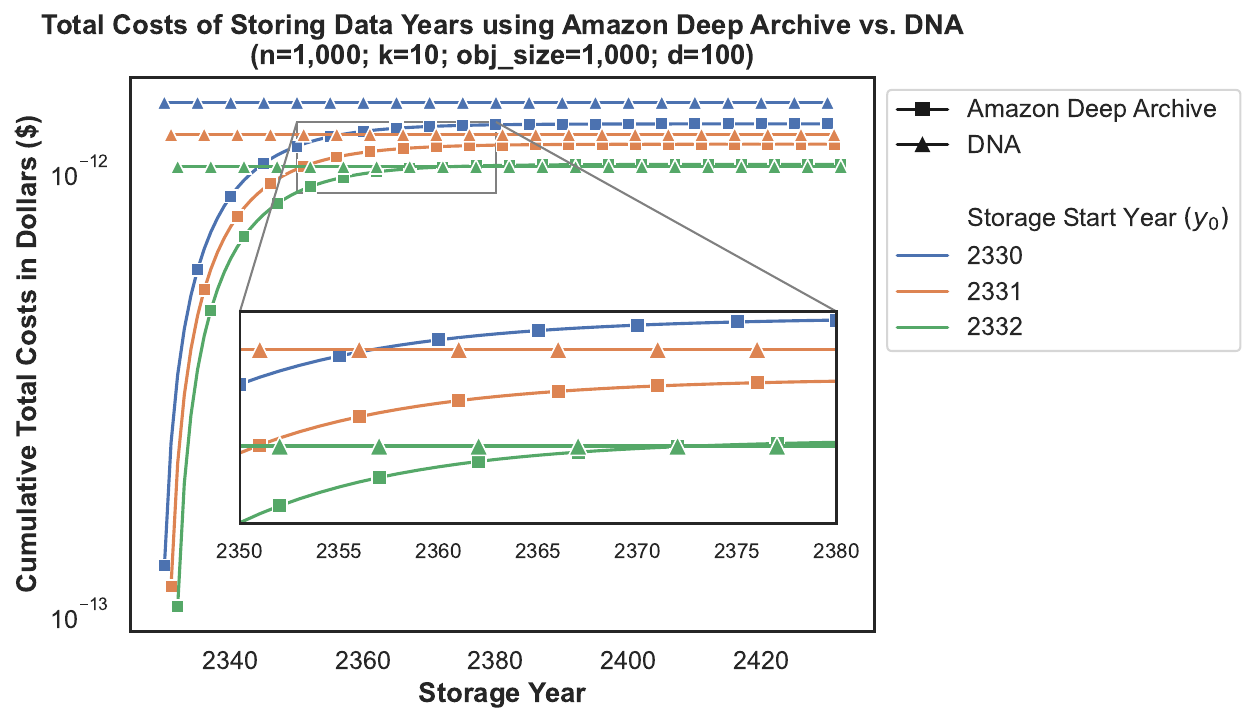}
		%\caption{The cumulative total cost trajectories for DNA storage vs. Amazon Deep Archive as a function of storage start year $y_0$.}
        \caption{}
		\label{h3.6}
	\end{subfigure}
    \caption{Comparing the total costs of using DNA vs. Amazon Deep Archive. \textbf{(a)} The total cost trajectories for DNA storage vs. Amazon Deep Archive. The point at which DNA first reaches cost parity with Amazon is marked, and two optimistic DNA cost-decline scenarios for the years 2035 and 2045 with their respective crossover years are also shown. \textbf{(b)} The cumulative total cost trajectories for DNA storage vs. Amazon Deep Archive as a function of storage start year $y_0$.}
\end{figure}

\autoref{h2.1} illustrates the total cost of DNA storage in comparison to Amazon Deep Archive, as a function of the storage start year ($y_0$). Since the costs of DNA synthesis and sequencing are expected to decline over time, the total cost of DNA storage decreases. Under the given parameters and current cost trends, DNA storage's total cost drop rate is $\approx 15.4\%$ per year, and it is projected to reach cost parity with Amazon Deep Archive by the year 2332. As shown in \textbf{Supplementary Figure 5}, cost parity is also projected within $\approx 10$ years of 2332 using Azure Deep Archive or tape on-premise.

\autoref{h2.1} also depicts the required DNA cost trajectories necessary to achieve parity with Amazon Deep Archive within the next 10 and 20 years, i.e., by 2035 and 2045, respectively. Notably, the cost gap in 2035 remains extremely large, with DNA storage being approximately eight orders of magnitude more expensive, matching the gap between DNA synthesis and tape write cost. Even by 2045, DNA storage is still projected to be about eight orders of magnitude more costly than Amazon Deep Archive.

\autoref{h3.6} plots the cumulative total cost of DNA storage compared to Amazon Deep Archive over the storage years as a function of storage start year ($y_0$). When $y_0=2332$, the two curves intersect in 2372, after which DNA remains the cheaper option. By setting $y_0=2331$, it is more cost-efficient to use DNA storage as compared to using tape one year earlier in $y_0=2330$.
%%
%#orange DNA vs blue tape = 2356
%# green x green = 2372
%%

The next series of experiments examines how the bit rate and oligonucleotide (oligo) length impact storage efficiency. Today's DNA synthesis technologies can only reliably produce relatively short oligos, ranging from 100 to 300 bases \cite{Hoose2023, 10.1145/3723166}. Additionally, every oligo reserves a fixed primer section, a short sequence used to selectively access and read the corresponding oligo using Polymerase Chain Reaction (PCR) \cite{https://doi.org/10.1002/adma.202307499}. Due to biochemical restrictions, only a few orthogonal primers exist, leading to multiple oligos sharing the identical primer sequences. To distinguish among them and reassemble the original data object, every oligo also carries an additional short address, commonly referred to as ``index''. Notably, when the oligo size is small, a larger portion of each oligo is dedicated to the primer and index (address section of the oligo), making the data payload section small. In contrast, larger oligos can store more data, increasing the overall bit rate of the DNA and potentially decreasing synthesis costs as less DNA is synthesised.

\begin{figure}[H]
	\centering
    \begin{subfigure}[t]{0.48\textwidth}
		\centering
		\includegraphics[width=\textwidth]{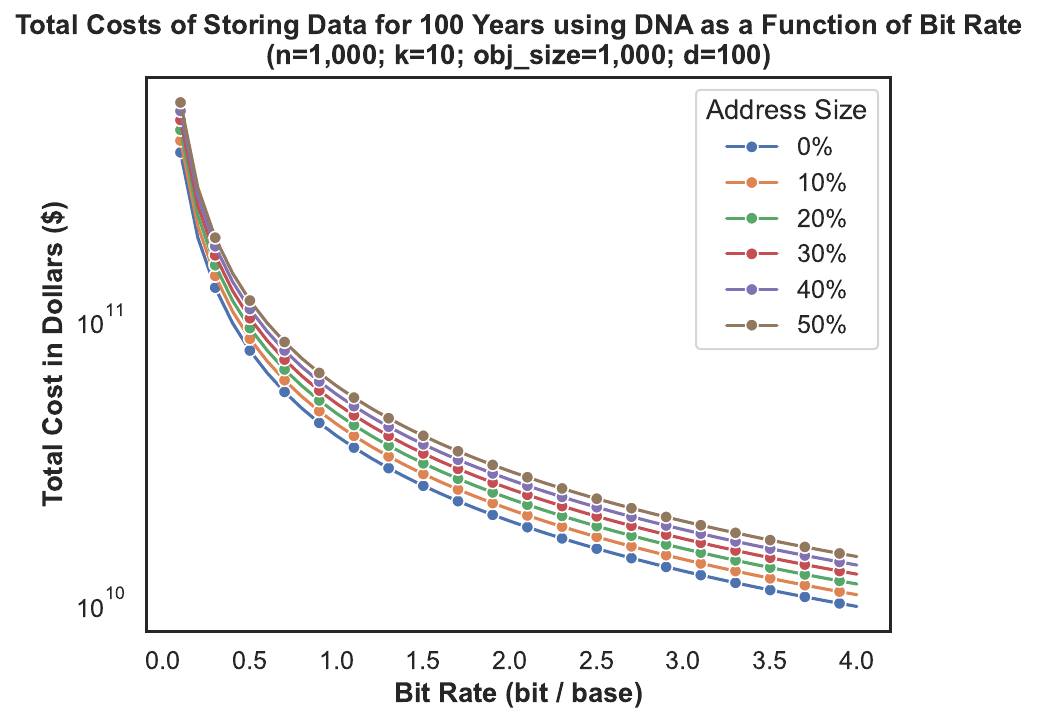}
		%\caption{Total Costs of using DNA storage with default parameters as a function of bit rate and address size.}
        \caption{}
		\label{h3.2}
	\end{subfigure}
	\hspace{1em}%\hfill
	\begin{subfigure}[t]{0.48\textwidth}
		\centering
		\includegraphics[width=\textwidth]{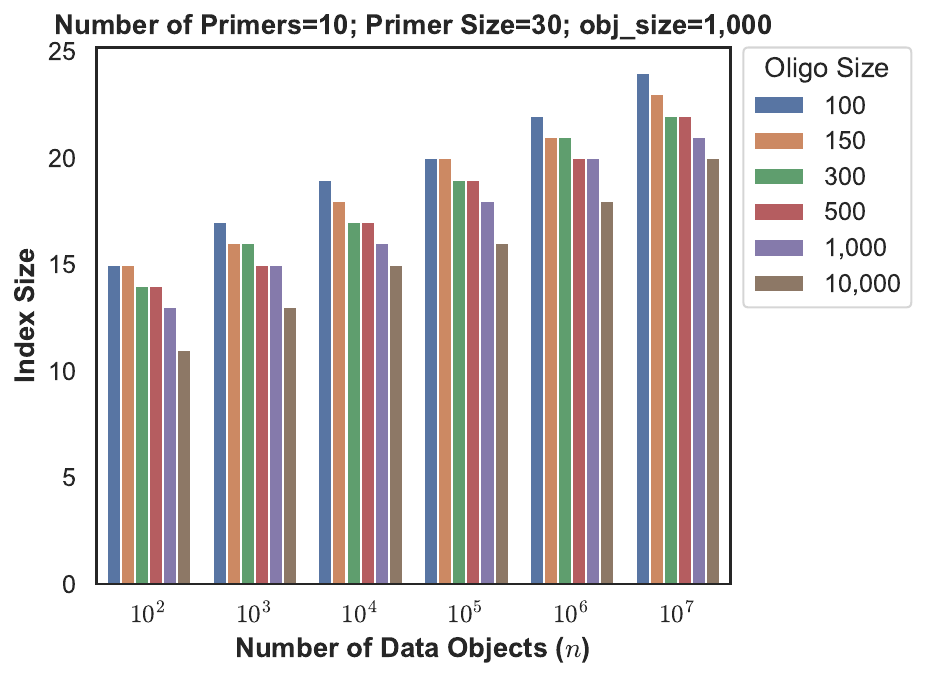}
		%\caption{The minimal required index size as a function of oligo size.}
        \caption{}
		\label{h.thomas.2}
	\end{subfigure}
	\hspace{1em}%\hfill
	\begin{subfigure}[t]{0.48\textwidth}
		\centering
		\includegraphics[width=\textwidth]{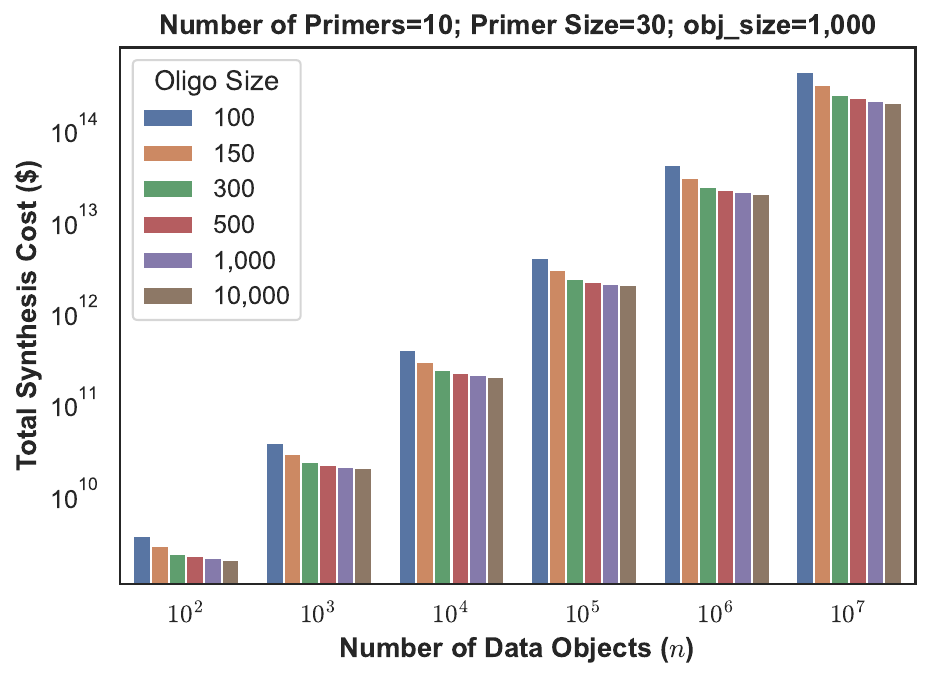}
		%\caption{The total cost of using DNA storage as a function of oligo size.}
        \caption{}
		\label{h.thomas.1}
	\end{subfigure}
    \caption{Examination of how the bit rate and address length (primer + index) influence the total cost of DNA‐based data storage. \textbf{(a)} Total Costs of using DNA storage with default parameters as a function of bit rate and address size. \textbf{(b)} The minimal required index size as a function of oligo size. \textbf{(c)} The total cost of using DNA storage as a function of oligo size.}
\end{figure}

\autoref{h3.2} depicts the total cost of using DNA storage with the default parameters as a function of bit rate and address size (primer plus index). \hl{Notably, our cost model is parametrised by the effective information density (bits per synthesised base). This quantity is intended to be net payload, i.e., after accounting for all overhead required for robustness: addressing/indexing, outer-code redundancy, and error correction needed to tolerate synthesis, storage (e.g., degradation), and sequencing errors, as well as dropout/coverage variability. Therefore, the economic impact of errors, error-correction and redundancy requirements are already captured implicitly in \mbox{\autoref{h3.2}}: higher error rates require stronger redundancy/error-correcting codes (ECC), which reduces the effective bits/base and moves the system left on the x-axis, increasing the total cost per stored bit accordingly.}

When the address occupies 50\% of the total DNA, then each oligo reserves half of its size for its address, halving the available data payload. Although reducing the address-to-oligo ratio does lower overall costs, it still fails to bring DNA storage in line with traditional storage media. Moreover, current archival pipelines compress data \cite{10.1117/12.825361}, increasing effective storage density; however, even raising the bit rate to an optimistic four bits per base does not reduce costs sufficiently to make DNA storage competitive with today's storage archive solutions.

Furthermore, \autoref{h.thomas.2} plots the minimal index size required given the number of available primers, number of data objects ($n$), and oligo size. For instance, larger oligos yield a smaller number of total oligos, as each oligo can carry more information. Similarly, when the oligo size is relatively small, more oligos encode the same data, requiring more index information to distinguish. For example, for an oligo size of $100$, and ten available primers (each 30 bases in size), the required index size is $15$ bases for indexing $n=100$ data objects, resulting in a total of 45 bases for the address. In contrast, setting the oligo size to $10{,}000$ and leaving the remaining parameters the same, the required index size is $11$, only four bases less. Notably, the index size grows logarithmically with the number of data objects and the oligo size if $\mathit{obj\_size}$ is constant. \textbf{Supplementary Figures 6-7} show the required index size for a different setting of the number of available primers and their sizes.

Moreover, \autoref{h.thomas.1} illustrates the resulting DNA synthesis costs as a function of oligo size and number of data objects, accounting for the necessary index for each oligo.
%While the cost drops with increasing oligo sizes, the obtained cost is still substantially higher than in traditional storage systems.
\hl{The reduction in synthesis cost with increasing oligo length in \mbox{\autoref{h.thomas.1}} arises because longer oligos (i) reduce the number of oligos required to encode a fixed file size and (ii) reduce the fractional indexing/addressing overhead needed per payload bit (see \mbox{\autoref{h.thomas.2}}). Each oligo must carry an index (and associated ECC/primer/structure overhead) so it can be mapped back to its position during decoding. When oligos are longer, this overhead is amortised over more payload bases, increasing the effective payload per synthesised base and lowering total synthesis required.}

% In addition, using current technologies, synthesising longer oligos is still less cost-efficient than shorter ones, resulting in higher overall costs \cite{10.1145/3723166}.

%\begin{figure}[H]
%	\centering
%	\begin{subfigure}[t]{0.85\textwidth}
%		\centering
%		\includegraphics[width=\textwidth]{figures/h3.3.pdf}
%		\caption{caption1}
%		\label{h3.3}
%	\end{subfigure}
%	\hspace{1em}%\hfill
%	\begin{subfigure}[t]{0.85\textwidth}
%		\centering
%		\includegraphics[width=\textwidth]{figures/h3.4.pdf}
%		\caption{caption2}
%		\label{h3.4}
%	\end{subfigure}
%   \hspace{1em}%\hfill
%	\begin{subfigure}[t]{0.85\textwidth}
%		\centering
%		\includegraphics[width=\textwidth]{figures/h3.5.pdf}
%		\caption{caption3}
%		\label{h3.5}
%	\end{subfigure}
%	\caption{Shared caption}
%	\label{h3.3+4+5}
%\end{figure}

% \begin{figure}[H]
% 	\centering
% 	\begin{subfigure}[t]{0.8\textwidth}
% 		\centering
% 		\includegraphics[width=\textwidth]{figures/h3_u_.7.pdf}
% 		\caption{}
% 		\label{h3.7}
% 	\end{subfigure}
% 	\hspace{1em}%\hfill
% 	\begin{subfigure}[t]{0.8\textwidth}
% 		\centering
% 		\includegraphics[width=\textwidth]{figures/h3_u_.8.pdf}
% 		\caption{}
% 		\label{h3.8}
% 	\end{subfigure}
%     \hspace{1em}%\hfill
% 	\begin{subfigure}[t]{0.8\textwidth}
% 		\centering
% 		\includegraphics[width=\textwidth]{figures/h3_u_.9.pdf}
% 		\caption{}
% 		\label{h3.9}
% 	\end{subfigure}
% 	\caption{Examining the cost of DNA storage if the initial write cost is set to zero for DNA and Amazon Deep Storage.}
% 	\label{h3.7+8+9}
% \end{figure}
The final series of experiments in \textbf{Supplementary Figure 8} examines the resulting costs of using DNA storage compared to Amazon Deep Archive, assuming zero initial write costs, and charging only for replacement writes plus maintenance and $k$ annual reads. In this scenario, setting $k=10$ in \textbf{Supplementary Figure 8a}, DNA storage is more cost-efficient than tape by $y_0=2050$, and the gap widens for larger $y_0$. In fact, our calculations reveal that DNA storage is more cost-efficient from $y_0=2036$ onward in this scenario. \textbf{Supplementary Figures 8b-8c} present the same analysis with $k=100$ and $k=1{,}000$, respectively. Remarkably, even with $k=100$, DNA storage deployed in $y_0=2050$ undercuts tape storage started in $y_0=2075$. These findings underscore the critical need for continued research into DNA synthesis, particularly efforts to drive down its cost if DNA-based storage is to become a truly economically viable alternative.

\section{Methods}\label{sec11}
\subsection{Cost Model}
To estimate the resulting cost of utilising various storage technologies, we propose a cost model to quantify the total expenditure involved in data storage over a specified period. This cost model incorporates three primary cost components: write costs, read costs, and running/maintenance costs. Additionally, it includes the necessity of periodic medium replacement and associated data migration, as determined by the durability characteristics of each storage medium.% Our model assumes that the storage system consists of data objects of fixed size, making the model directly applicable to object stores, commonly used in archival storage systems. However, it can be extended to reflect more complex data storage systems.

%\subsection{Model Parameters}
\autoref{tab:params} introduces the various parameters used with corresponding descriptions.
\begin{table}[t]
\centering
\begin{tabular}{ll}
\toprule
\textbf{Parameter} & \textbf{Description} \\
\midrule
$n$ & Number of objects to store \\
$\mathit{obj\_size}$ & Size of each object (in megabytes, MB) \\
$d$ & Total duration of storage (in years) \\
$k$ & Number of objects read per year \\
$L$ & Lifetime (in years) of the storage medium \\
$M$ & Number of media migrations (computed from $L$ and $d$)\\
$y_0$ & The first year in which storage starts\\
$C_{\text{write}}(y)$ & Write cost (in U.S. dollars) of 1 MB in year $y$ \\
$C_{\text{read}}(y)$ & Read cost (in U.S. dollars) of 1 MB in year $y$ \\
$C_{\text{run}}(y)$ & Running and maintenance costs (in U.S. dollars) of $1$ MB in year $y$ \\
\bottomrule
\end{tabular}
\caption{Parameters used in the storage cost model.}\label{tab:params}
\end{table}
Following the table above, $n$ objects are stored for $d$ years, where every object is of fixed size $\mathit{obj\_size}$. Thus, a total of $ n \cdot \mathit{obj\_size}$ MB of data are stored in the system. To account for read access, $k$ objects are read every year, i.e., $k \cdot \mathit{obj\_size}$ MB of data are read every year. The parameter $y_0$ denotes the first year in which data is stored onward. The functions $C_{\text{write}}(y)$, $C_{\text{read}}(y)$, and $C_{\text{run}}(y)$ return the cost of writing, reading, and maintaining a megabyte of data in year $y$, respectively.

Furthermore, the storage medium's lifetime $L$ serves as a flexible parameter representing media durability in years. If desired, $L$ could be expressed as a function $L(y)$ to reflect changes in lifetime in year $y$. However, we treat it as a constant for simplicity. Hence, the total number of media replacements $M$ can be calculated as follows:
\begin{equation}\label{eq:M}
    M = \bigg\lfloor \frac{d-1}{L} \bigg\rfloor
\end{equation}
For example, using \autoref{eq:M} with $L=30$ and $d=100$, the total number of required media replacements is $M=\Big\lfloor \frac{100 - 1}{30} \Big\rfloor = \lfloor3.3\rfloor = 3$. For $y_0=2025$, the media is migrated in 2055, 2085, and 2115.
%Accordingly, estimating storage media replacement costs requires identifying the number of replacements and the specific years in which they are expected to occur. The following procedure in \autoref{algo:replacement_years} returns the set of years where the required replacements take place.
%\begin{algorithm}[H]
%\caption{Calculate Device Replacement Years}\label{algo:replacement_years}
%\begin{algorithmic}[1]
%\Function{Replacements\_Years}{$y_0$, $d$, $L$}
%    \State $\mathit{current\_year} \gets y_0$\label{algo:replacement_years:y0}
%    \State $\mathit{repl\_years} \gets \{\}$
%    \While{$\mathit{current\_year} < y_0 + d$}
%        \State $\mathit{lifetime} \gets \mathit{L}$ \label{algo:replacement_years:curr_year}
%        \State $\mathit{current\_year} \gets \mathit{current\_year} + \mathit{lifetime}$
%        \If{$\mathit{current\_year} < y_0 + d$}
%            \State $\mathit{repl\_years} \gets \mathit{repl\_years} \cup \{\mathit{current\_year}\}$
%        \EndIf
%    \EndWhile
%    \State \Return $\mathit{repl\_years}$
%\EndFunction
%\end{algorithmic}
%\end{algorithm}
%Starting from year $Y=y_0$, the algorithm iteratively adds $L(Y)$ to the current year, yielding the next replacement year. Whenever this new year still precedes the deadline $y_0 + d$, i.e., in which storage ends, it is appended to the replacement set. The loop terminates once the projected replacement falls beyond the end of the storage period.

Based on the parameters above, the resulting total write $C_{\text{write\_total}}$, read $C_{\text{read\_total}}$, and running $C_{\text{run\_total}}$ costs are calculated as:
%\begin{equation}\label{eq:model}
%\begin{aligned}
%C_{\text{total}} ={}& \underbrace{\sum_{t=0}^{M} C_{\text{write}}(y_0 + t\cdot L) \cdot (n \cdot %\mathit{obj\_size})}_{\text{Initial write for $t=0$, and migrations for $t>0$}} \\[0.5em]
%&+ \underbrace{\sum_{t=0}^{d-1} C_{\text{run}}(y_0 + t) \cdot (n \cdot %\mathit{obj\_size})}_{\text{Running cost over $d$ years}} \\[0.5em]
%&+ \underbrace{\sum_{t=0}^{d-1} C_{\text{read}}(y_0 + t) \cdot (k \cdot \mathit{obj\_size})}_{\text{Read %access cost over $d$ years}}
%\end{aligned}
%\end{equation}

\begin{equation}\label{eq:write_cost}
C_{\text{write\_total}} =
\left( n \cdot \mathit{obj\_size} \right) \cdot \!\!\!\!\!\!\!\!\!\!\!\!\!\!
\underbrace{\sum_{t=0}^{M} C_{\text{write}}(y_0 + t\cdot L)}_{\text{Initial write ($t=0$) + migrations ($t>0$)}}
\end{equation}

\begin{equation}\label{eq:read_cost}
C_{\text{read\_total}} =
\left( k \cdot \mathit{obj\_size} \right) \cdot \!\!\!\!\!\!\!
\underbrace{\sum_{t=0}^{d-1} C_{\text{read}}(y_0 + t)}_{\text{Read access cost over $d$ years}}
\end{equation}

\begin{equation}\label{eq:run_cost}
C_{\text{run\_total}} =
\left( n \cdot \mathit{obj\_size} \right) \cdot \!\!\!\!\!
\underbrace{\sum_{t=0}^{d-1} C_{\text{run}}(y_0 + t)}_{\text{Running cost over $d$ years}}
\end{equation}

In \autoref{eq:write_cost}, the write cost $C_{\text{run\_total}}$ consists of the initial write cost in year $y_0$, and $M$ subsequent migrations for one MB of data. The write cost function $C_{\text{write}}$ is multiplied by $n \cdot \mathit{obj\_size}$ to return the cost for the entire data volume. For each migration or media replacement, the same $n$ data objects are rewritten to new storage media. While a dedicated migration cost function could be defined, we assume $C_{\text{write}}$ also applies to future migrations.

Similarly, the total read cost $C_{\text{read\_total}}$ in \autoref{eq:read_cost} is computed by multiplying the sum of reading one MB of data over the storage period by the read data size $k \cdot \mathit{obj\_size}$.

Lastly, \autoref{eq:run_cost} calculates the total running cost as the sum of annual operating costs over the storage period, multiplied by the total data volume. Finally, summing the total write, read, and run costs yields the total storage costs: $C_{\text{total}} = C_{\text{write\_total}} + C_{\text{run\_total}} + C_{\text{read\_total}}$.

\subsection{Estimating Cost Functions}
The following sections estimate the cost functions for each storage method: DNA storage, Amazon S3 Glacier Deep Archive (Amazon Deep Archive), Azure Blob Archive, and tape on-premise storage.
\subsubsection{DNA Storage Costs}
To capture the drop in DNA synthesis and sequencing costs over the past years, we fitted an exponential decay model to the annual cost per base. To compute the cost for one megabyte of data (1 megabyte of data contains $4 \cdot 10^6$ DNA bases), the following exponential cost function is used:
\begin{equation}\label{eq:dna_read_cost}
    C_\text{write$\mid$read}(y) \;=\; C_{t_0} \cdot e^{-\lambda (y - t_0)} \cdot 4 \cdot {10}^6
\end{equation}
where $t_0$ is the reference year, $\lambda$ is the cost decay rate, and $C_{t_0}$ is set to the write or read cost of one DNA base in the reference year $y=t_0$, respectively.

\begin{figure}[H]
	\centering
	\begin{subfigure}[t]{0.47\textwidth}
		\centering
		\includegraphics[width=\textwidth]{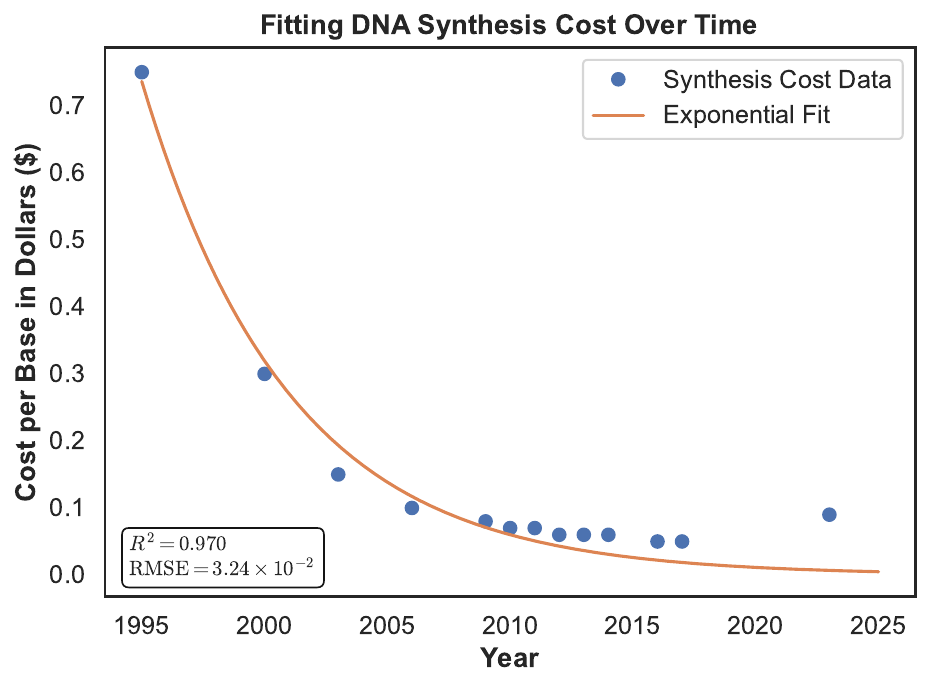}
        \caption{}
		\label{fig:synthesis_cost}
	\end{subfigure}
	\hspace{1em}%\hfill
	\begin{subfigure}[t]{0.47\textwidth}
		\centering
		\includegraphics[width=\textwidth]{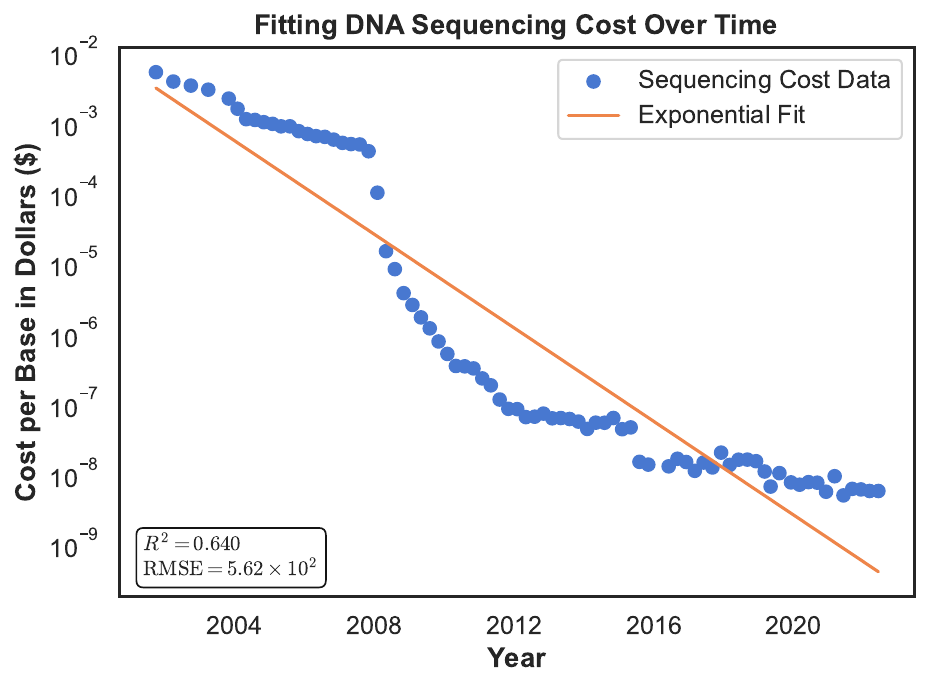}
        \caption{}
		\label{fig:sequencing_cost}
	\end{subfigure}
    \caption{DNA synthesis and sequencing cost data fit to their exponential curves, and their projections into the future.}\label{fig:exponentials}
\end{figure}

\hl{\mbox{\autoref{fig:exponentials}} presents exponential cost models for DNA synthesis and sequencing obtained by fitting curves to historical data from \mbox{\cite{carlson2022dna, Carr2009, kosuri2014large, wetterstrand2023dnacosts, lin2022enable}}. The models were estimated using least-squares regression: synthesis costs were fitted via non-linear least squares on the original scale, whereas sequencing costs were modelled using linear regression applied to logarithmically transformed data. In both cases, model performance was evaluated on the original scale using the coefficient of determination ($R^{2}$) and the root mean square error (RMSE). Here, $R^{2}$ indicates the proportion of variance explained by the model, while RMSE quantifies the typical deviation between predicted and observed costs. The resulting models are subsequently used as time-dependent cost functions for writing to and reading using DNA.}

Furthermore, using \autoref{eq:dna_read_cost} from above, we derived the following DNA synthesis cost function's parameters:
\[(C_{t_0}=0.3193,\; t_0=2000, \; \lambda=0.1671)\]

For example, plugging in these parameters into \autoref{eq:dna_read_cost}, the cost for writing one MB of data in $2025$ can be calculated as:
\begin{alignat}{1}
    C_\text{write}(2025) \;&=\; 0.3193 \cdot e^{-0.1671 \cdot (2025 - 2000)} \cdot 4 \cdot {10}^6\\
    &\approx 19{,}588\; \text{U.S. dollars}
\end{alignat}

Similarly, the resulting DNA sequencing cost function uses the following parameters:
\[(C_{t_0}=0.0087,\; t_0=2000, \; \lambda=0.4791)\]

Using the read cost function above, reading one MB of data in $2025$ can be calculated as:
\begin{alignat}{1}
    C_\text{read}(2025) \;&=\; 0.0087 \cdot e^{-0.4791 \cdot (2025 - 2000)} \cdot 4 \cdot {10}^6\\
    &\approx 0.22\; \text{U.S. dollars}
\end{alignat}

Notably, writing DNA is around five orders of magnitude more expensive than reading DNA.

Finally, we set the running and maintenance costs for using DNA to $C_{\text{run}}(y) = 0$ as these costs are negligible. \hl{While DNA archives will incur operational costs (e.g., environmental control, security, periodic inventory/integrity checks, and facility overhead), in our baseline, we treat these as insignificant relative to the dominant cost drivers: DNA synthesis for writing and sequencing for reading. That is because (i) properly dried/encapsulated DNA can be stored at low temperature with very low energy input, (ii) access events (and therefore integrity checks) are infrequent for cold archival workloads, and (iii) even conservative facility costs amortised over high-density media are orders of magnitude smaller on a per-GB-year basis than current write/read costs. The intent of this simplification is not to claim that the operational cost is exactly zero, but that it is second-order in the regimes we analyse.}

%This is justified because DNA synthesis and sequencing inherently involve handling DNA, and storing the resulting DNA incurs minimal cost due to its insignificant space requirements and low energy consumption.

\subsubsection{Amazon Deep Archive Costs}
The costs of Amazon Deep Archive are taken from the publicly advertised rates \cite{aws2025}. Write and read operations incur both a fixed per‑request fee and a variable per-megabyte data transfer fee. To capture expected price declines over time, we apply an adjustable exponential decay factor to all cost components.

We model the costs for Amazon S3 Glacier Deep Archive using mid-2025 US East (N. Virginia) pay-as-you-go prices. Let $y$ denote the current year, $\mathit{obj\_size}$ the object size in MB, and $r$ the annual price drop rate. The write, read, and running costs per MB in year $y$ are:
\begin{alignat}{1}
C_{\text{write}}(y) &= \frac{p_{\text{put}}}{\mathit{obj\_size}} \cdot (1-r)^{y-2025}\label{eq:amazon:write}\\
C_{\text{read}}(y) &= \left( \frac{p_{\text{req}}}{\mathit{obj\_size}} + p_{\text{data}} \right) \cdot (1-r)^{y-2025}
\label{eq:amazon:read}\\
C_{\text{run}}(y) &= p_{\text{maint}} \cdot (1-r)^{y-2025}\label{eq:amazon:run}
\end{alignat}
where $r$ is the cost decline rate per year. The operations \textit{put} and \textit{request} are used to charge $p_{\text{put}}$, and $p_{\text{req}}$ for a single write and read operation, respectively. Additionally, a fee of $p_{\text{data}}$ is charged per MB retrieved. Finally, $p_{\text{maint}}$ is paid for every MB stored for maintaining the required storage capacity for one year. According to the current costs of using Amazon Deep Archive, these values are set as follows:
\begin{equation}
p_{\text{put}} = \frac{0.05}{1000},\:
p_{\text{req}} = \frac{0.025}{1000},\:
p_{\text{data}} = \frac{0.0025}{1024},\:
p_{\text{maint}} = \frac{0.00099 \cdot 12}{1024}
\end{equation}

In \autoref{eq:amazon:write} and \autoref{eq:amazon:read}, the costs are calculated by dividing $p_{\text{put}}$ and $p_{\text{req}}$ by $\mathit{obj\_size}$ because our cost model in the main manuscript multiplies the associated cost functions by the the number of objects and object size, i.e., $n \cdot \mathit{obj\_size}$, whereas Amazon charges per operation for \textit{put} or \textit{req}, and not per MB.

\subsubsection{Azure Blob Archive Costs}
Similar to Amazon Deep Archive, the costs of Azure Blob Archive are taken from the publicly advertised rates \cite{azure2025}. Like Amazon Deep Archive, write and read operations incur a fixed per-request fee and a variable per-megabyte data transfer fee. We apply an adjustable exponential decay factor to all cost components to capture expected price declines over time.

We model the costs for Azure Blob Archive using mid-2025 East US~2 pay-as-you-go prices for flat namespace accounts. Based on the current advertised costs, \autoref{eq:amazon:write}, \autoref{eq:amazon:read}, and \autoref{eq:amazon:run} are paramertised with:

%Let $y$ denote the current year, $\mathit{obj\_size}$ the object size in MB, and $r$ the annual price drop rate. The write, read, and running costs per MB in year $y$ are:
%\begin{alignat}{1}
%C_{\text{write}}(y) &= \frac{p_{\text{put}}}{\mathit{obj\_size}} \cdot (1-r)^{y-2025} %\label{eq:azure:write}\\
%C_{\text{read}}(y) &= \left( \frac{p_{\text{req}}}{\mathit{obj\_size}} + p_{\text{data}} \right) \cdot (1-%r)^{y-2025} \label{eq:azure:read}\\
%C_{\text{run}}(y) &= p_{\text{maint}} \cdot (1-r)^{y-2025} \label{eq:azure:run}
%\end{alignat}
%where $r$ is the cost decline rate per year. Here, $p_{\text{put}}$ and $p_{\text{req}}$ denote the fees for a single write and read operation, respectively, while $p_{\text{data}}$ is the fee per MB retrieved.
%Finally, $p_{\text{maint}}$ is the cost per MB to store data for one year.
%Based on the current price offerings, the parameters are set to the following values:
\begin{equation}
p_{\text{put}} = \frac{0.10}{10\,000},\:
p_{\text{req}} = \frac{5}{10\,000},\:
p_{\text{data}} = \frac{0.02}{1024},\:
p_{\text{maint}} = \frac{0.002 \cdot 12}{1024}
\end{equation}

%In \autoref{eq:azure:write} and \autoref{eq:azure:read}, we divide $p_{\text{put}}$ and $p_{\text{req}}$ by %$\mathit{obj\_size}$ because the main cost model multiplies the per-MB cost functions with $n \cdot %\mathit{obj\_size}$, whereas Azure charges per operation for \textit{put} and \textit{req}, and not per MB.

\subsubsection{Tape On-premise Costs}
Tape costs are modelled as per-MB charges without per-request fees, and are based on the analysis in \cite{fujifilm2025tcotool}. We apply an adjustable exponential decay to capture expected price improvements over time.

Let $y$ denote the current year and $r$ is the cost decline rate per year. The write, read, and running costs per MB in year $y$ are:
\begin{alignat}{1}
C_{\text{write}}(y) &= 6.39 \cdot 10^{-6} \cdot (1-r)^{y-2025} \label{eq:tape:write}\\
C_{\text{read}}(y)  &= 0 \label{eq:tape:read}\\
C_{\text{run}}(y)   &= 6.71 \cdot 10^{-6} \cdot (1-r)^{y-2025} \label{eq:tape:run}
\end{alignat}
where $C_{\text{read}}(y)$ is zero, as the associated costs are accounted for in the running cost.

\section{Discussion}\label{sec12}

This work evaluates the economic feasibility and broader implications of implementing DNA data storage systems, guided by the four hypotheses \textbf{H1-H4} mentioned in the introduction. Our results offer validation and essential considerations for the potential of DNA storage systems.

Hypothesis \textbf{H1} posited that the cost of DNA synthesis presents the primary barrier preventing widespread adoption of DNA storage today. While DNA synthesis and, especially, DNA sequencing costs have been declining dramatically over the past two decades, they still incur several orders of magnitude higher costs than traditional storage technologies, such as tape. Notably, the cost of DNA sequencing has been declining by several orders of magnitude in just a few years, thanks to the broad adoption of sequencing technologies across research and commercial sectors \cite{Mardis2017}, significant technological advancements like next-generation sequencing that enable massively parallel sequencing \cite{Shendure2008}, and increasing clinical demand for genomic information in medical diagnostics \cite{Manolio2013}. While the cost of reading tape is around four orders of magnitude cheaper than DNA sequencing today, the cost of DNA synthesis remains ten orders of magnitude higher than writing tape. However, as with sequencing, DNA synthesis could also undergo disruptive technological breakthroughs and significant cost reductions in the future. This discrepancy currently significantly amplifies the initial write cost of storing data in DNA, even under read-intensive scenarios, confirming \textbf{H1}.

Furthermore, addressing \textbf{H2}, following historical cost trends in DNA synthesis and sequencing, they are projected to reach cost parity with tape only by 2332, i.e., more than 300 years into the future. However, omitting the initial write costs for both DNA and tape, cost parity is achieved by 2036. Additionally, our analysis indicates that the storage duration has a negligible impact on the overall cost. Because DNA is extraordinarily durable, data migration is only necessary after very long intervals, and by that time, synthesis and sequencing technologies have become exponentially cheaper. As a result, the expense of periodically rewriting data onto newly synthesised DNA is negligible. Hence, unlike in traditional storage systems, where long-term retention can drive increased costs, in DNA storage, the dominant factor remains the upfront synthesis cost rather than the duration of preservation. These findings underscore the importance of accelerating investment and advancement of DNA synthesis technologies to make DNA storage an economically viable alternative, thereby verifying hypothesis \textbf{H2}.

Moreover, improvements in DNA storage density result in limited cost declines. Recall that DNA synthesis and sequencing are multiple orders of magnitude more expensive than using tape. For instance, if the data were twice as densely packed on DNA, i.e., half of the DNA is required, then DNA synthesis and sequencing costs only drop by a factor of two, validating \textbf{H3}.

%%%%%
Furthermore, integrating DNA as a storage medium within data centres introduces unique architectural and economic considerations. DNA storage relies on encoding information into fixed-length DNA oligos, a constraint that directly influences both costs and retrieval strategies. Larger data objects are distributed among more oligos than smaller data objects. Each oligo contains a unique addressing section consisting of two parts: (i) a primer for PCR amplification, which enables retrieval, and (ii) an index that helps recover the corresponding segment of the data object. While the same primer can be used for all oligos of a given data object, the index size increases with the data object size, as shown in our results. Consequently, larger data objects require more addressing space on each oligo, increasing the total DNA needed and the associated cost.

Moreover, only a limited number of primers are available per DNA pool due to biochemical constraints \cite{10.1145/3723166, 10.1093/nargab/lqab126}. When the number of encoded data objects exceeds the number of available primers, then multiple data objects must share the same primer. To further distinguish oligos with the same primer but encoding different data objects, additional information is inserted into the index section. This, however, causes retrieval of irrelevant oligos when reading a single data object, which must then be filtered out, amplifying bandwidth and data recovery time.

To mitigate this limitation, physical separation of DNA oligos becomes necessary, a strategy that optimises read cost and latency, and ensures fine-grained accessibility of distinct data objects. However, this physical separation increases space requirements and thereby data management overhead, incurring extra costs. Thus, further research is required to calibrate access granularity and develop practical strategies that balance storage space efficiency with retrieval performance, especially in the context of large-scale data centre deployment.

Looking ahead, the future of DNA storage will be shaped not only by advancements in molecular technologies but also by improvements in automation and scalability. Automation of DNA synthesis and sequencing workflows through microfluidics, enzymatic synthesis platforms, and high-throughput laboratory systems has the potential to reduce operational costs and increase throughput significantly \cite{DENG2024100222, Hoose2023, Jo2024, Ma2024, Pichon2024, yu2024high}. As these technologies mature, economies of scale could further drive down costs, enabling more widespread adoption beyond niche archival use cases. Integration with existing data centre infrastructure, development of standardised interfaces, and advances in random access capabilities will be critical to transitioning DNA storage from a research concept to a practical component of the storage ecosystem \cite{10.1145/3723166, Zhou2024}. Ultimately, the combination of technical innovation and industrial automation may be key to unlocking the full potential of DNA as a viable medium for long-term, high-density data storage.

Finally, despite the extraordinarily high costs of DNA storage, its use for preserving critical or culturally significant information, such as governmental archives, foundational scientific datasets, cultural heritage collections, or records essential for recovery after global crises, may still be justified. In such cases, the unmatched longevity, density, and stability of DNA could offer distinct advantages over traditional storage technologies that outweigh purely economic considerations. Importantly, we are living in a period of unprecedented data generation, with global data volumes growing exponentially and at an ever-increasing pace. At no point in history have we produced such vast quantities of digital information, nor have we had the means to preserve them reliably across centuries or millennia. While significant challenges remain, such as mitigating the risk of a digital dark age \cite{kuny1998digital} or ensuring long-term interpretability \cite{Gervasio2024}, DNA storage presents a compelling solution for safeguarding information far beyond the lifespan of existing media. These findings support \textbf{H4}, confirming that the unique characteristics of DNA storage open up application domains that are fundamentally inaccessible to traditional storage technologies.

%\section{Conclusion}\label{sec13}

\backmatter

\bmhead{Supplementary information}
Supplementary Figures are available.

%\bmhead{Acknowledgements}

%Acknowledgements are not compulsory. Where included they should be brief. Grant or contribution numbers may be acknowledged.

\section*{Declarations}

%Some journals require declarations to be submitted in a standardised format. Please check the Instructions for Authors of the journal to which you are submitting to see if you need to complete this section. If yes, your manuscript must contain the following sections under the heading `Declarations':

\begin{itemize}
\item Funding: The work has been supported by the EiC Pathfinder projects NEO (101115317) and DNAMIC (101115389) within Horizon Europe.
\item Conflict of interest/Competing interests: 
The authors declare no competing interests.
\item Ethics approval and consent to participate: Not applicable
\item Consent for publication: Not applicable
\item Data availability: 
Our code generates all the necessary data to reproduce our results. Our code is available at: \url{https://github.com/alexelshaikh/Economic\_DNA}
\item Author contribution: 
A.E.S. and T.H. conceptualised the work. A.E.S. wrote the manuscript. T.H. and B.S. contributed to the design of the experiments, analysis, revision, and preparation of the manuscript.
\end{itemize}

\noindent
%If any of the sections are not relevant to your manuscript, please include the heading and write `Not applicable' for that section. 

\bigskip

%\bigskip\noindent
%Springer journals and proceedings: \url{https://www.springer.com/gp/editorial-policies}

%\bigskip\noindent
%Nature Portfolio journals: \url{https://www.nature.com/nature-research/editorial-policies}

%\bigskip\noindent
%\textit{Scientific Reports}: \url{https://www.nature.com/srep/journal-policies/editorial-policies}

%\bigskip\noindent
%BMC journals: \url{https://www.biomedcentral.com/getpublished/editorial-policies}
%\end{flushleft}

\begin{appendices}




\end{appendices}


\bibliography{references}% common bib file

\end{document}